\documentclass[prl,aps,twocolumn,showpacs]{revtex4}
\usepackage{graphicx,amssymb,amsmath,color,psfrag}
\usepackage{amsthm}
\usepackage{amsfonts}
\usepackage{algorithmic}
\usepackage{enumerate}
\usepackage{latexsym}

\input{tcilatex}
\begin{document}

\title{Goos-H\"{a}nchen Shifts of Partially Coherent Light Fields}
\author{Li-Gang Wang$^{1,2,3}$, Shi-Yao Zhu$^{4}$, and M. Suhail Zubairy$%
^{1,4}$}
\affiliation{$^{1}$Institute for Quantum Science and Engineering (IQSE) and Department of
Physics and Astronomy, Texas A$\&$M University, College Station, TX
77843-4242, USA\\
$^{2}$Department of Physics, Zhejiang University, Hangzhou 310027, China\\
$^{3}$The National Center for Mathematics and Physics, KACST, P. O. Box
6086, Riyadh 11442, Saudi Arabia\\
$^{4}$Beijing Computational Science Research Center, Beijing, 100084, China}

\begin{abstract}
We investigate the Goos-H\"{a}nchen (GH) shifts of partially coherent fields
(PCFs) by using the theory of coherence. We derive a formal expression for
the GH shifts of PCFs in terms of Mercer's expansion, and then clearly
demonstrate the dependence of the GH shift of each mode of PCFs on spatial
coherence and beam width. We discuss the effect of spatial coherence on the
resultant GH shifts, especially for the cases near the critical angles, such
as totally reflection angle.
\end{abstract}

\date{\today }
\maketitle

Goos-H\"{a}nchen (GH) Shift refers to a tiny (lateral) displacement, from
the path expected from geometrical optics, upon total reflection \cite%
{GoosHanceh1947}. This effect has been extended into other fields that
involve the coherent-wave phenomena, such as neutron waves \cite%
{Igonatovich2004,Haan2010}, electron waves \cite{Beenakker2009,Wu2011}, and
spin waves \cite{Dadoenkova2012}. It was explained by Artmann \cite%
{Artman1948} that the different transverse wave vectors of a light beam
undergo different phase changes and sum of these waves form a reflected beam
with a lateral shift. Recently, it was shown \cite{Chen2012} that the GH
shift is the sum of Renard's conventional energy flux plus a
self-interference shift. The self-interference shift originates from the
interference between the incident and the reflected beams. Furthermore, it
was discovered that the classical Fresnel formulas for laws of refraction
and reflection of light are not applicable to partially coherent light \cite%
{Lahiri2012}. These explanations indicate that the interference or coherence
of light is very important to the GH shift.

In 2008, we numerically showed the effect of spatial coherence on the change
of the GH shift near the critical angles \cite{WANGLG2008}. Later, an
experiment \cite{Schwefel2008} showed the large difference between the
measured GH shift of a partially coherent LED light and the theoretical
result of a coherent light. However, the very recent investigations \cite%
{Aiello2011,Merano2012,Loffler2012,WANGLG2012,WANGLG2013} have raised an
important issue \textquotedblleft whether the spatial coherence of the
partially coherent fields (PCFs) influences the GH
shifts?\textquotedblright\ Although the exact numerical results, calculated
from our previous theory \cite{WANGLG2008}, are in good agreement with the
experimental data \cite{Merano2012,Loffler2012,WANGLG2013}, it is necessary
to reconsider this issue thoroughly and bring to light the role of spatial
coherence on the GH shift.

In this Letter, we use the exact theory of coherence to investigate the GH
shift of PCFs. First we derive a formal expression to calculate the GH shift
of PCFs in terms of the mode expansion of PCFs. Based on this expression, we
explain the physical mechanism about the dependence of the GH shift on the
spatial coherence and the beam width. Finally, we suggest a proposal for
showing the new effect of the spatial coherence on the practical GH shift
below the critical angles. 

First we derive the GH shift of PCFs based on the coherence theory \cite%
{Mandel1995}. For the two-dimensional PCFs, one usually uses the
cross-spectral density (CSD), $W(x_{1},z_{1};x_{2},z_{2},\nu )$, to describe
its propagation, where $(x_{1},z_{1})$ and $(x_{2},z_{2})$ are the
coordinates of the two points in the fields, and $\nu $ is the frequency of
light. For simplicity, we omit the symbol $\nu $. According to the theory of
coherence, $W(x_{1},z_{1};x_{2},z_{2})$ can be expressed in the form of
Mercer's expansion, namely \cite{Mandel1995}%
\begin{equation}
W(x_{1},z_{1};x_{2},z_{2})=\dsum\limits_{m}\beta _{m}\psi _{m}^{\ast
}(x_{1},z_{1})\psi _{m}(x_{2},z_{2}),  \label{Mercer1}
\end{equation}%
where $\psi _{m}$ are the eigenfunctions and $\beta _{m}\geq 0$ are the
corresponding eigenvalues. We rewrite it in another form,%
\begin{equation}
W(x_{1},z_{1};x_{2},z_{2})=\dsum\limits_{n}\beta
_{m}W^{(m)}(x_{1},z_{1};x_{2},z_{2}),  \label{WWWn}
\end{equation}%
where $W^{(m)}(x_{1},z_{1};x_{2},z_{2})=\psi _{m}^{\ast }(x_{1},z_{1})\psi
_{m}(x_{2},z_{2})$ represents the CSD of a field that is perfect coherent.
When PCFs are reflected at the interface ($z_{1}=z_{2}=z$) between two
media, each mode $\psi _{m}$ (for both TE and TM polarization) experiences a
GH shift. Therefore the reflected CSD for a single mode $\psi _{m}$, at the
interface, can be formally written as%
\begin{align}
& W_{r}^{(m)}(x_{1},z_{1};x_{2},z_{2})=W_{r}^{(m)}(x_{1},z;x_{2},z)  \notag
\\
& =|\overline{r}(\theta _{0},\delta \theta _{m})|^{2}\psi _{m}^{\ast
}(x_{1}-\Delta _{m},z)\psi _{m}(x_{2}-\Delta _{m},z),  \label{WRM}
\end{align}%
where $\delta \theta _{m}$ and $\Delta _{m}$ are the angular spread and the
practical GH shift of the $m$th mode, respectively, and $\overline{r}(\theta
_{0},\delta \theta _{m})$ is the averaged reflection coefficient within $%
\delta \theta _{m}$ around the incident angle $\theta _{0}$ for the $m$th
mode. Since $\delta \theta _{m}$ may be very broad for a large $m$, the
first-order Taylor expansion (FOTE) on the reflection coefficient $r$ around 
$\theta _{0}$ can fail \cite{Horowitz1971}. Thus $\Delta _{m}$ are very
different for different modes due to the size effect of each mode, and they
are also different from the prediction of the formulae $\Delta _{FOTE}=-%
\func{Re}[i\left. \frac{\partial \ln r}{\partial \theta }\right\vert
_{\theta \rightarrow \theta _{n}}]$ or $-\frac{\lambda d\phi _{r}}{2\pi
d\theta }$, which is based on the stationary phase method under the FOTE 
\cite{Artman1948,LiCF2003,Simon1989}, here $\phi _{r}$ is the phase of $r$.
Therefore, the total reflected CSD of a PCF at the interface is given by%
\begin{align}
& W_{r}(x_{1},z;x_{2},z)  \notag \\
& =\dsum\limits_{m}w_{m}(\theta _{0},\delta \theta _{m})\psi _{m}^{\ast
}(x_{1}-\Delta _{m},z)\psi _{m}(x_{2}-\Delta _{m},z),
\end{align}%
where $w_{m}(\theta _{0},\delta \theta _{m})=|\overline{r}(\theta
_{m},\delta \theta _{m})|^{2}\beta _{m}$ represents the weight of the $m$th
reflected mode. Then the intensity of the reflected beam is 
\begin{equation}
I_{r}(x,z)=\dsum\limits_{m}w_{m}(\theta _{0},\delta \theta _{m})|\psi
_{m}(x-\Delta _{m},z)|^{2}.  \label{Intensity}
\end{equation}%
Using the normalized first moment of the light field \cite%
{Bretenaker1992,Shadrivov2003}, $\Delta =\left. \dint xI_{r}(x,z)dx\right/
\dint I_{r}(x,z)dx$, we obtain the resultant GH shift as follows%
\begin{equation}
\Delta =\frac{\dsum\limits_{m}w_{m}(\theta _{0},\delta \theta _{m})\Delta
_{m}}{\dsum\limits_{m}w_{m}(\theta _{0},\delta \theta _{m})},
\label{NEWSHIFT}
\end{equation}%
where the normalization condition, $\int |\psi _{n}(x,z;\nu )|^{2}dx=1$, has
been used. Equation (\ref{NEWSHIFT}) is a formal expression for calculating
the practical GH shift of PCFs. This result is different from that in Refs. 
\cite{Aiello2011} and \cite{Loffler2012}. In Refs. \cite{Aiello2011} and 
\cite{Loffler2012}, since all shifts $\Delta _{m}$ are assumed to be $\Delta
_{FOTE}$, so that $\Delta =\Delta _{FOTE}$ is independent of spatial
coherence. However, this is not true for PCFs, especially for the incoherent
light fields. In the following discussion, we will see that, as $m$
increases, there is a large difference between $\Delta _{m}$ and $\Delta
_{FOTE}$. Even for a coherent beam, $\Delta _{m}$ also changes due to the
finite-size effect of practical light beams \cite{Lai1986,Golla2011}. Thus
the exact expression for $\Delta _{m}$ for each mode should be defined as 
\cite{Bretenaker1992,Shadrivov2003} 
\begin{equation}
\Delta _{m}=\left. \dint x\left\vert \psi _{m}^{r}(x,z)\right\vert
^{2}dx\right/ \dint \left\vert \psi _{m}^{r}(x,z)\right\vert ^{2}dx,
\label{DeltaNN}
\end{equation}%
where $\psi _{m}^{r}$ is the reflected field of the $m$th mode at the
interface. Therefore, for an incoherent light field, we must include the
contributions of the shifts $\Delta _{m}$ of all modes to the resultant GH
shift $\Delta $.

Next we consider how/why the spatial coherence affects the GH shift of each
mode of PCFs. We briefly review a famous example: Gaussian Shell-model (GSM)
beam, which is an excellent model for describing PCFs \cite{Mandel1995}. The
normalized eigenfunctions and eigenvalues of GSM beams are given by \cite%
{Mandel1995} (also see Refs. \cite{Gori1980,Starikov1982})%
\begin{equation}
\psi _{m}(x)=(\frac{2c}{\pi })^{1/4}\frac{1}{(2^{m}m!)^{1/2}}%
H_{m}[x(2c)^{1/2}]e^{-cx^{2}},  \label{psin}
\end{equation}%
and $\beta _{m}=A^{2}(\frac{\pi }{a+b+c})^{1/2}(\frac{b}{a+b+c})^{m}$, where 
$H_{m}(x)$ are the Hermite polynomials, $a,b$ and $c$ are positive
quantities and are defined as: $a=(4\sigma _{s}^{2})^{-1}$, $b=(2\sigma
_{g}^{2})^{-1}$, $c=[a^{2}+2ab]^{1/2}.$ Here $\sigma _{s}$ and $\sigma _{g}$
are the beam half-width and the spectral coherence width of PCFs,
respectively. The ratio of the eigenvalue $\beta _{m}$ to the lowest
eigenvalue $\beta _{0}$ is evidently given by \cite{Mandel1995} 
\begin{equation}
\frac{\beta _{m}}{\beta _{0}}=\left[ \frac{1}{%
(q^{2}/2)+1+q[(q/2)^{2}+1]^{1/2}}\right] ^{m},  \label{RATIOPCF}
\end{equation}%
where $q=\sigma _{g}/\sigma _{s}$ is a measure of the degree of global
coherence of a GSM source. Obviously, for $q\gg 1$ ($\sigma _{g}\gg \sigma
_{s}$), $\beta _{m}/\beta _{0}\approx q^{-2m}$. This implies that, for all $%
m>0$, $\beta _{m}\ll \beta _{0}$ and hence the behavior of the beam is well
approximated by the lowest-order mode. However, for $q\ll 1$ ($\sigma
_{g}\ll \sigma _{s}$), $\beta _{m}/\beta _{0}\approx 1-mq$. Thus, for a very
incoherent light, a large number of modes (of the order $1/q$)\ are needed
to represent the light field adequately.

\begin{figure}[tbp]
\includegraphics[width=8cm]{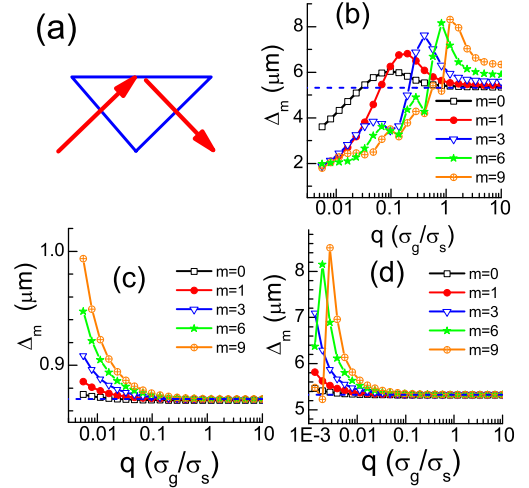}
\caption{(Color) (a) The schematic of total reflection from a prism. (b-d)
The dependence of the GH shifts $\Delta _{m}$ of each mode on the spatial
cohernce ($q$) at different values of $\protect\theta _{0}$: (b,d) $\protect%
\theta _{0}=41.5^{\circ }$ and (c) $\protect\theta _{0}=45^{\circ }$. In (b
and c) $\protect\sigma _{s}=0.1$ mm, and in (d) $\protect\sigma _{s}=2$ mm.
The blue dashed lines in (b, c, and d) denote the values of $\Delta _{FOTE}$%
. }
\label{Fig:fig1}
\end{figure}

Since each mode of PCFs is perfectly coherent, it is easy to obtain the GH
shift for each mode under a certain incident angle upon total internal
reflection, as illustrated in Fig. 1(a). Here we use the coherent
angular-spectral theory \cite{Shadrivov2003,LiCF2003,soboleva2012,Lahiri2012}%
. From Eq. (\ref{psin}), we readily obtain its angular spectrum, $\tilde{\psi%
}_{m}(k_{x})$, by using a Fourier transformation. For an inclined case with $%
\theta _{0}>0$, $\tilde{\psi}_{m}(k_{x})$ becomes $\tilde{\psi}%
_{m}(k_{x}-k_{x0})$ with the replacement $\sigma _{s}\rightarrow $ $\sigma
_{s}\sec \theta _{0}$ and $\sigma _{g}\rightarrow $ $\sigma _{g}\sec \theta
_{0}$, where $k_{x}$ is the transverse component of the wavenumber $k$ of
light in the first medium, and $k_{x0}=k\sin \theta _{0}$. Therefore the
reflected field of the $m$th mode is given by 
\begin{equation}
\psi _{m}^{r}(x)=\frac{1}{\sqrt{2\pi }}\int r(k_{x})\tilde{\psi}%
_{m}(k_{x}-k_{x0})\exp [ik_{x}x]dk_{x}.  \label{phsirm}
\end{equation}%
Then, using Eq. (\ref{DeltaNN}), we can obtain all shifts $\Delta _{m}$ in
any situation. In the following calculations, we take the refractive index
of the prism $n=1.514$ at wavelength $\lambda =675$ nm, so the critical
angle of the totally internal reflection is $\theta _{c}=41.34^{\circ }$.
Here we only present the result for the TM polarization, due to the
similarity between TM and TE polarizations.

\emph{Effect of spatial coherence}.---Figures 1(b) and 1(c) show the typical
dependence of the GH shifts $\Delta _{m}$ of the $m$th mode on the spatial
coherence ($q$) under different values $\theta _{0}$: (b) $\theta
_{0}=41.5^{\circ }$ and (c) $\theta _{0}=45^{\circ }$. In these two cases,
we take $\sigma _{s}=0.1$ mm ($>>\lambda $). From Figs. 1(b) and 1(c), it is
found that, near the critical angle $\theta _{c}$, the absolute shifts $%
\Delta _{m}$ are strongly dependent on $q$. For $m=0$, the value $\Delta
_{0} $ slightly increases when $q$ is gradually close to $0.1$, and then it
decreases as $q$ further decreases. As $m$ increases, the changes $\Delta
_{m}$ become more dramatic with the decreasing $q$, and more oscillations
appear due to the fact that the part components of $\widetilde{\psi }%
_{m}(k_{x}-k_{x0})$ have been cut off below $\theta _{c}$ as $\widetilde{%
\psi }_{m}(k_{x}-k_{x0})$ is broadened with the decreasing of $q$. From Fig.
1(c), for the cases of $\theta _{0}$ being far away from $\theta _{c}$, the
values $\Delta _{m}$ change much more for larger $m$. Thus it is expected
that there must be a difference between the coherent and incoherent limits 
\cite{WANGLG2008,WANGLG2012,WANGLG2013}. Comparing Fig. 1(b) with Fig. 1(c),
it is also found that the changes of $\Delta _{m}$ near $\theta _{c}$ are
more remarkable than that for the cases being away from $\theta _{c}$.

In Fig. 1(d), we plot another situation for the dependence of $\Delta _{m}$
on $q$ at $\theta _{0}=41.5^{\circ }$, with $\sigma _{s}=2$ mm. Although $%
\theta _{0}$ is still near to $\theta _{c}$, the changes of $\Delta _{m}$ in
Fig. 1(d) are considerably small for $q\geq 0.01$. This is due to the effect
of beam width $(2\sigma _{s})$\ on $\Delta _{m}$ discussed below. From Fig.
1(d), it is clear that there is a large difference between $\Delta _{m}$ and 
$\Delta _{FOTE}$ in the incoherent limit ($q<0.01$). When $m$ increases,
some oscillations may also appear for a sufficient small $q$.

\begin{figure}[tbp]
\includegraphics[width=8cm]{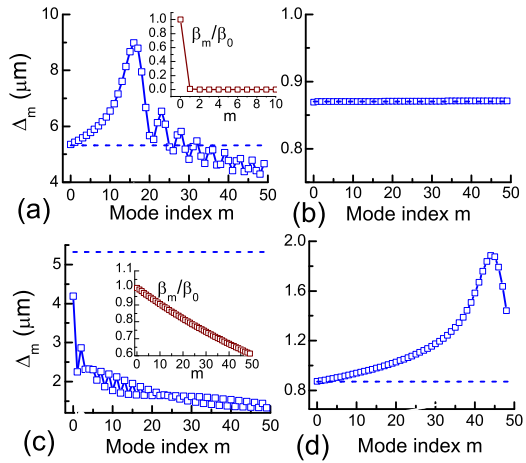}
\caption{(Color) The dependence of the GH shifts $\Delta _{m}$ of each mode
on the mode index $m$ under two limits (a, b) $q=10$ and (c, d) $q=0.01$
with incident angles (a, c) $\protect\theta _{0}=41.5^{\circ }$ and (b, d) $%
\protect\theta _{0}=45^{\circ }$. Insets in (a) and (c) show the value of $%
\protect\beta _{m}/\protect\beta _{0}$ as a function of $m$ for $q=10$ and $%
q=0.01$, respectively. The dashed lines in (a-d) denote the values of $%
\Delta _{FOTE}$.}
\label{Fig:fig2}
\end{figure}

In Fig. 2, we further show the changes of $\Delta _{m}$ as a function of $m$
under two limits: (1) $q=10$ and (2) $q=0.01$ with (a, c) $\theta
_{0}=41.5^{\circ }$ and (b, d) $\theta _{0}=45^{\circ }$. Insets in Figs. 2
(a, c) show the value of $\beta _{m}/\beta _{0}$ as a function of $m$ for $%
q=10$ and $q=0.01$, respectively. For the fully coherent limit ($q\gg 1$),
when $\theta _{0}$ is close to $\theta _{c}$ [see Fig. 2(a)], $\Delta _{m}$
vary dramatically as $m$ increases; while when $\theta _{0}$ is far away
from $\theta _{c}$ [see Fig. 2(b)], $\Delta _{m}$ are nearly independent of $%
m$ and they are overlapped with the corresponding value $\Delta _{FOTE}$.
Thus, in the full-coherent limit, $\Delta _{m}$ are independent of $m$ only
under the cases of $\theta _{0}$ being far away from $\theta _{c}$.
Meanwhile it is only the shifts $\Delta _{0}$ of the lowest mode ($m=0$)
that mainly contribute to the resultant GH shift $\Delta $ since $\beta _{m}$
do decrease quickly for $m>0$, see the inset in Fig. 2(a). For the
completely incoherent limit ($q\ll 1$), see Figs. 2(c, d), whether $\theta
_{0}$ is close to or far away from $\theta _{c}$, $\Delta _{m}$ do vary as $%
m $ increases; and the contributions of the higher-order modes must be
included since $\beta _{m}$ changes very slowly for $m>0$, see the inset in
Fig. 2(c). This leads to the resultant GH shift $\Delta $ deviated from the
full-coherent limit.

\emph{Effect of beam width }$(2\sigma _{s})$.--- We note that the beam width
of the PCFs plays a role on the GH shift, since the effective width (2$%
\sigma _{m}^{eff}$) of each mode $\psi _{m}$ is related to both $\sigma _{s}$
and $\sigma _{g}$ \cite{Mandel1995}. From Eq. (\ref{psin}), we can obtain $%
\sigma _{m}^{eff}=\sqrt{2m+1}\sigma _{s}/[1+(4/q^{2})]^{1/4}$ and its
corresponding angular spread $\delta \theta _{m}=\frac{180\sqrt{2m+1}}{\pi
k\sigma _{s}}[1+(4/q^{2})]^{1/4}$ (in the unit of degree). In order to keep
the fixed values, $\sigma _{m}^{eff}$ and $\delta \theta _{m}$, if $\sigma
_{s}$ increases, the value of $q$ must decrease. In other words, for a fixed
value of $q$, if $\sigma _{s}$ increases, then $\sigma _{m}^{eff}$ increases
but $\delta \theta _{m}$ decreases. This means that increasing $\sigma _{s}$
suppresses the effect of spatial coherence ($q$) on the GH shift. By
comparing Fig. 1(b) and Fig. 1(d), it is clear that, increasing $\sigma _{s}$
leads to the weakening of the effect of spatial coherence on the GH shift.
It should be pointed out that, for a coherent beam, the effect of beam width
has been investigated in the very early literature \cite%
{Horowitz1971,Lai1986} and has also been experimental demonstrated \cite%
{Cowan1977,Bretenaker1992}. Therefore it is expected that the beam width has
also an effect on the GH shift for PCFs.

\begin{figure}[tbp]
\includegraphics[width=8cm]{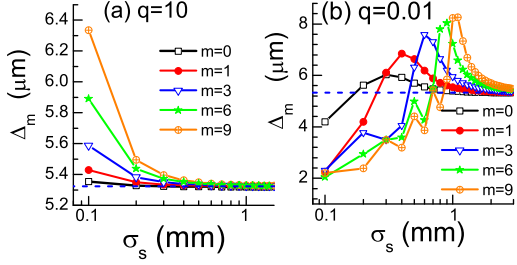}
\caption{(Color) Effect of beam half-width on the GH shifts $\Delta _{m}$ of
each mode with different fixed $q$: (a) $q=10$, and (b) $q=0.01$, with $%
\protect\theta _{0}=41.5^{\circ }$.}
\label{Fig:fig3}
\end{figure}

Figure 3 shows the detailed effect of $\sigma _{s}$ on the GH shift $\Delta
_{m}$ near $\theta _{c}$. From Fig. 3(a), even for a full-coherent limit
with $q=10$, when $\sigma _{s}$ is small enough (\TEXTsymbol{<}$0.3$ mm),
the values $\Delta _{m}$ begin to be significantly different from the value
of $\Delta _{FOTE}$, and such a difference becomes larger as $m$ increases.
Remember that it is only the lowest mode ($m=0$) that dominates the
resultant GH shift $\Delta $ in the full-coherent limit, thus other $\Delta
_{m}$ with $m>0$ have no contributions to $\Delta $. However, in the
incoherent limit ($q=0.01$), see Figs. 3(b), when $\sigma _{s}$ is larger
than $2$ mm, the difference between $\Delta _{m}$ and $\Delta _{FOTE}$
gradually disappears due to the suppressing effect of $\sigma _{s}$ on $%
\Delta _{m}$; while for the cases when $\sigma _{s}$ is smaller than $2$ mm
in our cases, $\Delta _{m}$ change dramatically and they are very different
from $\Delta _{FOTE}$.

In fact, on comparing Fig. 3(b) with Fig. 1(b), we find that the role of $%
\sigma _{s}$ on $\Delta _{m}$ for a small $q$ is similar to the role of $q$
on $\Delta _{m}$ for a small $\sigma _{s}$. On comparing Fig. 3(a) with Fig.
1(d), we can also find that the role of $\sigma _{s}$ on $\Delta _{m}$ for a
large $q$ is similar to the role of $q$ on $\Delta _{m}$ for a large $\sigma
_{s}$. Therefore, both $\sigma _{s}$ and $q$ have the equivalent role on the
GH shift.

Now we have known the roles of $\sigma _{s}$ and $q$ on the GH shift of each
mode of PCFs, and have explained why/how they affect the shift $\Delta $.
However, it is inconvenient for using Eq. (\ref{NEWSHIFT}) to obtain $\Delta 
$ since it is time-consuming to know all the practical shifts $\Delta _{m}$
for PCFs when $q$ is very small. For example, if $q=0.01,$ we need $100$.
modes at least. There is a much realistic method to directly obtain $\Delta $%
. Based on our previous investigation, the exact expression for the
intensity of the reflected PCFs, at the interface of two media ($%
z_{1}=z_{2}=0$), can be given by \cite{WANGLG2008}, 
\begin{eqnarray}
I_{r}(x,0) &=&W_{r}(x,0;x,0)  \notag \\
&=&\frac{1}{2\pi }\dint \dint r^{\ast
}(k_{x1})r(k_{x2})W_{i}(k_{x1},0;k_{x2},0)  \notag \\
&&\times \exp [-i(k_{x1}-k_{x2})x]dk_{x1}dk_{x2},  \label{WWWREXACT}
\end{eqnarray}%
where $W_{i}(k_{x1},0;k_{x2},0)$ is the initial CSD in the spatial
angular-frequency domain at $z=0$, and $r(k_{x2})$ is the reflection
coefficient. Substituting Eq. (\ref{WWWREXACT}) into the definition of $%
\Delta $, we can obtain the GH shift of PCFs at $z=0$ by the exact numerical
method.

\begin{figure}[tbp]
\includegraphics[width=7cm]{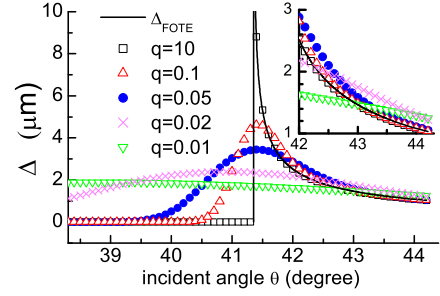}
\caption{(Color) The resultant GH shifts as a function of incident angle
under different values of $q$ with a fixed value of $\protect\sigma _{s}=0.2$
mm. Insets are the details above $\protect\theta _{c}$.}
\label{Fig:fig4}
\end{figure}

Finally, let us briefly discuss how to experimentally demonstrate the effect
of spatial coherence on the GH shift of PCFs, since the recent experiments 
\cite{Merano2012,Loffler2012} have not revealed this effect. From the above
discussion, we have already seen that the large value $\sigma _{s}$ weakens
the effect; and near $\theta _{c}$, the spatial coherence has a larger
effect. Thus one should take a small $\sigma _{s}$ and measure the absolute
GH shift $\Delta $ of PCFs near $\theta _{c}$ in the experiment. In Fig. 4,
we predict a dependence of its spatial coherence on the absolute GH shifts
for experimental reference. In this case, we take $\sigma _{s}=0.2$ mm, and
consider five cases: $q=10$, $0.1$, $0.05$, $0.02$ and $0.01$. From Fig. 4,
we see that, for a full coherent light, there are non-zero GH shifts above $%
\theta _{c}$ but zero below $\theta _{c}$, and all the shifts for $\Delta $
are overlapped with the curves of $\Delta _{FOTE}$. However, for a PCF or an
incoherent light field, the GH shifts above $\theta _{c}$ may be smaller or
larger than $\Delta _{FOTE}$, see the insets in Fig. 4. More importantly,
the GH shifts below $\theta _{c}$ are no longer equal to zero. This is a
distinct result for PCFs, which is completely different from the
full-coherent prediction. Actually, the latter effect has been observed in a
recent experiment \cite{Schwefel2008}, where the authors observed a non-zero
GH shift below the critical angle, but they cannot explain it. The non-zero
GH shifts of PCFs below $\theta _{c}$ are very similar to the effect of the
narrow beam width of the coherent beam on the GH shifts \cite%
{Lai1986,Gray2007}. Since our curves in Fig. 4 have the same property with
other experiments \cite{Bretenaker1992,Yin2004,Unterhinninghofen2011}, we
hope our suggestions could lead to a direct experimental observation of this
effect in the system of the total internal reflection.

In summary, we have found that both the spatial coherence and beam width of
PCFs have strong effect on its GH shift, which are explained by the formal
equation (\ref{NEWSHIFT}) by using the exact theory of coherence. Our
results show that the spatial coherence of PCFs play an important role to
determine the resultant GH shifts. Finally, we suggest a potential
experiment to demonstrate this effect and display a distinct effect for
experimental verification. These effects are very important to the
applications of the GH shift in nano- or micro-scaled structures \cite%
{Schwefel2008,Yin2004}, where light beams are usually focused into the small
region and the coherence may play an important role. Our results are also
important to the applications of the GH effect in other fields, such as
neutron systems \cite{Igonatovich2004,Haan2010} and electronic systems \cite%
{Beenakker2009,Wu2011}, where the coherent sources are usually not available.

\begin{acknowledgments}
This research is supported by NPRP grant 4-346-1-061 by the Qatar National
Research Fund (QNRF) and a grant from King Abdulaziz City for Science and
Technology (KACST). This work is also supported by NSFC grants (No.
61078021, 11174026 and No. 11274275), and the grants by the National Basic
Research Program of China (No. 2012CB921602,3 and 2011CB922203).
\end{acknowledgments}

\end{document}